\documentclass[aip,apl,twocolumn,showpacs,reprint]{revtex4-1}
\usepackage{graphicx}
\usepackage{dcolumn}
\usepackage{bm}
\usepackage{amssymb}
\newcommand{\beq}{\begin{equation}}  
\newcommand{\eeq}{\end{equation}}  
\newcommand{\beqa}{\begin{eqnarray}}  
\newcommand{\eeqa}{\end{eqnarray}}

\begin{document}

\title{Intrinsic spin noise in MgO magnetic tunnel junctions.}

\author{ F. Delgado}
\affiliation{International Iberian Nanotechnology Laboratory (INL),
Av. Mestre Jos\'e Veiga, 4715-330 Braga, Portugal}
\author{K. Lopez}
\affiliation{Department of Mechanical Engineering,  Massachusetts Institute of Technology,   Cambridge, MA 02139 USA }
\affiliation{Mechanical Engineering Department, Stanford University, Stanford, CA 94305 USA}
\author{ R. Ferreira}
\affiliation{International Iberian Nanotechnology Laboratory (INL),
Av. Mestre Jos\'e Veiga, 4715-330 Braga, Portugal}
\author{ J. Fern\'andez-Rossier}
\affiliation{International Iberian Nanotechnology Laboratory (INL),
Av. Mestre Jos\'e Veiga, 4715-330 Braga, Portugal}
\affiliation{Departamento de F\'isica Aplicada, Universidad de Alicante, 03690 San Vicente del Raspeig, Spain
}

\begin{abstract}
%

%
We consider two intrinsic sources of  noise in ultra-sensitive magnetic field sensors based on MgO  magnetic tunnel junctions, coming both from
$^{25}$Mg nuclear spins ($I=5/2$, 10$\%$ natural abundance), and $S=1$ Mg-vacancies.
While nuclear spins induce noise peaked in the MHz frequency range,  the vacancies noise peaks  in the GHz range. We find that the nuclear noise in submicron devices has a similar magnitude than  the $1/f$ noise, while the vacancy-induced noise  dominates in the GHz range. Interestingly,  the noise spectrum under a finite magnetic field gradient  may provide spatial information about the spins in the MgO layer.

 \end{abstract}

\maketitle

Magnetic tunnel junctions (MTJ) with ferromagnetic electrodes and a MgO tunnel barrier have a very large room temperature tunneling magnetoresistance (TMR).\cite{Parkin_Kaiser_natmat_2004,Yuasa_Nagahama_natmat_2004}  As a result, they are widely used  for magnetic sensing applications where room-temperature ultra-high sensitivity, circuit integration and low fabrication cost are essential. 
Engineering of multilayer MTJ devices has allowed building devices whose resistance scales linearly with the applied magnetic field.
 If this linear relation holds at arbitrarily small field,  the devices can operate as sensors for  magnetic fields as small as permitted by the different sources of noise.  Broadly speaking, these can be classified in two groups, electric and magnetic.\cite{Ingvarsson_Xiao_prl_2000,Parkin_Jiang_ieee_2003,Klaassen_Xing_ieee_2005,Freitas_Ferreira_jphysc_2007,Egelhoff_Pong_sensors_2009,Lei_Li_magnetics_2011}  The former includes shot-noise, Johnson-Nyquist noise,  electric $1/f$ noise or noise due to charge trapping in the oxide barrier. The second includes fluctuations in the magnetic orientation of the electrodes due to collective precessional modes, $1/f$ magnetic noise, domain wall motion and so on.

MgO based TMR sensors with an area of $1\mu {\rm m}^2$ feature sensitivities of up to $pT/\sqrt{{\rm Hz}}$ 
limited by white noise background.\cite{Freitas_Ferreira_jphysc_2007,Egelhoff_Pong_sensors_2009}
 This striking sensitivity leads us to address the following intriguing question: to which degree the magnetic field created by spins 
in the subnanometer thick MgO barrier can be a source of noise that limits the performance of these devices?  Or reversing the terms of the question:  could the electrical noise of a MgO-MTJ  probe the spin noise of the barrier?   

\begin{figure}
\includegraphics[width=1.\linewidth,angle=0]{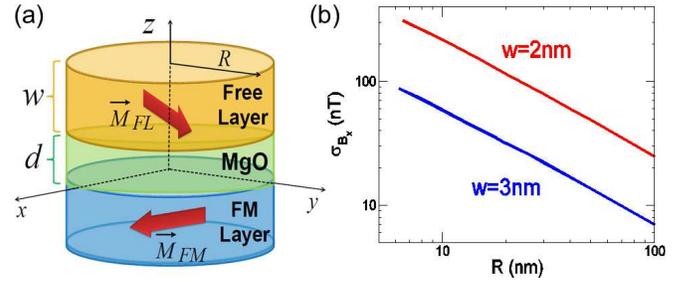}
\caption{ \label{fig1} (color online) (a) Scheme of a MTJ sensing device. (b)
Variation of the standard deviation of the average  field in the free layer 
with the detector radius $R$ for a device with $d=0.5$ nm. 
}
\end{figure}

The MgO barrier certainly hosts   
the only stable Mg spinful nuclear isotopes, $^{25}$Mg ,with   nuclear spin, $I=5/2$.  Thus, taking into account that the lattice constant of the MgO is $4.212$ \AA{}, and its natural relative abundance 
of 10$\%$,\cite{Berglund_Wieser_pachem_2011}
the volumetric density of nuclear spins is $\rho_m=1.32$ spins/nm$^3$. 
 The maximal magnetic field created by one of such nuclear spins, at  a distance $l$,
 reaches   $4.3{\rm nm}^3/l^3$ $\mu$T.   In addition,  the MgO barrier hosts  a density of Mg vacancies\cite{Halliburton_Kappers_prl_1973,Halliburton_Cowan_prb_1973,Rose_Halliburton_jphys_1974,Araujo_Kapilashrami_apl_2010}  which have electronic spin $S= 1$, each of which will create a magnetic field 3 orders of magnitude larger. 

In standard MTJ sensing devices, one magnetic layer is designed to have its magnetization pinned by exchange coupling to an antiferromagnet while the other is free to rotate, see Fig. \ref{fig1}(a).\cite{Freitas_Ferreira_jphysc_2007,Parkin_Jiang_ieee_2003}
Then, the relevant figure of merit is  given by the sum of all nuclear fields, averaged over the entire free layer (FL) sensing electrode
\begin{equation}
 \vec{{\cal B}}\left(\vec{m}_i\right)\equiv
\frac{1}{V}\int_V \vec{B}\left[\vec{m}_i\right](\vec{r}) dV,
\label{average}
\end{equation}
where the integral is over the volume $V$ of the detector and 
$\vec{B}\left[\vec{m}_i\right](\vec{r})$ corresponds to the magnetic field created at position ${\vec r}$ by the set of magnetic dipoles $\left\{ {\vec m}_i\right\}$. 
If all the nuclear spins were fully polarized, they would create an average field that,  for a cylindrical device with $R=100 nm$, would lead to ${\cal B}_{max}\sim 0.1\; \mu$T, which motivates a detailed study of the  nuclear spin noise in this system.

At room temperature the average nuclear spin orientation is vanishingly small, and so it is the average magnetic field they create, but statistical fluctuations of the nuclear spin orientation create magnetic noise.  For the calculation of its  statistical properties,
the following relation between the average sensing layer field, Eq.  (1), and  the nuclear
magnetic moments $\vec{m}_i$  is extremely useful:
\begin{equation}
{\cal B}_a 
= \sum_{i,b} \Xi_{ab}(i) 
m_b(i),
\label{B3}
\end{equation}
where 
\begin{equation}
 \Xi_{ab}(i) = \frac{\mu_0}{4\pi} \frac{1}{V}  \int_V dV 
\frac{ n_b(i)  n_a(i) -\delta_{ab}}{|\vec{r}-\vec{r}_i|^3}.
\label{B30}
\end{equation}
$\Xi_{ab}(i)$ is a geometrical factor that relates the $a$ component of the average detector field
to the  $b$  component  of the nuclear magnetic moment $i$, with $a,b=x,y,z$ . 
The linear relation in Eq. (\ref{B30})  permits relating the quantum  statistical properties of the nuclear spins to those of the sensing layer average in a straightforward way,  in particular if one assumes that  different  nuclear spins are uncorrelated.
In this way, the standard deviation of the $a$-magnetic field component created by the fully randomized nuclear spins, 
defined as $\sigma_{{\cal B}_a}^2\equiv \left(\langle {\cal B}_a - \langle{\cal B}_a\rangle\right)^2$, where the brackets stand for the quantum statistical average, 
can be written as
\beqa
\sigma_{ {\cal  B}_a}^2= \left(g^{*}\mu_N \right)^2 I(I+1)\sum_{i,b} \Xi_{ab}(i)^2,
\label{total-noise}
\eeqa
 where we have used $\langle m^2\rangle=g^{*2} \mu_N^2 I(I+1)$, with $\mu_N$ the nuclear magneton and $g^*$ the effective g-factor ($g^*\approx0.342$ for the $^{25}$Mg).\cite{Stone_atdat_2005}

The quantity $\sigma_{ {\cal  B}_a} $ represents the $a$-component of the nuclear  magnetic field  noise integrated over the entire frequency range. 
In addition, if the nuclear spins are randomized,  we will  find that in cylindrical devices like the one in Fig.~\ref{fig1}(a), 
  $\sigma_{{\cal  B}_x}= \sigma_{{\cal  B}_y} $.  Since we can safely neglect changes in the magnitude of the magnetization, the nuclear noise field can only be efficient in rotating the FL magnetization, which by design of these sensors, can only happen in the plane of the layer. Therefore, only the noise along the in-plane direction $x$ perpendicular to the equilibrium magnetization, will compromise the sensor accuracy. 
Figure \ref{fig1}(b) shows the numerically calculated
  $\sigma_{{\cal  B}_x}$ for two devices with FL thickness $w=2$ and 3$nm$, 
and barrier thickness  $d=0.5$nm   a function of $R$.  
Positions $\vec{r}_i$ in the MgO layer have been randomly chosen and we have checked that results do not significantly depend on the random distribution.
From Fig. \ref{fig1}, we can extrapolate and get that  for $R=1\;\mu$m and $w=3$nm, $\sigma_{{\cal B}_x}\approx 10$nT.

 From our numerics, we find that   $\sigma_{{\cal B}_x}$ grows linearly with $1/R$ except for very small devices $R\lesssim 10$nm.  Thus, the relevance of the nuclear spin noise increases for smaller sensors.  Notice that  from Eq. (\ref{total-noise}) it is ostensible that $\sigma_{ {\cal  B}_a}^2$ scales proportionally  to  $N$, the number of nuclear  spins in the barrier.  This is a consequence of the linear relation in Eq. (\ref{B3}) on one hand, and the linear scaling between the statistical fluctuations of the {\em total magnetic moment} and the number of spins.   
 \cite{Sleator_Hahn_prl_1985,Degen_Poggio_prl_2007} Nevertheless, in our case the $1/R$ scaling of the standard deviation of the  magnetic field comes from  the scaling of integral (\ref{B30}).

In addition to the unavoidable nuclear spin noise,   MgO can have a certain density  of oxygen and magnesium 
vacancies.\cite{Wertz_Auzins_dfs_1959,Halliburton_Kappers_prl_1973,Halliburton_Cowan_prb_1973,Rose_Halliburton_jphys_1974}
 The most likely spinfull vacancy  in MgO are the Mg vacancies, $V_{Mg}$, with concentrations that vary between $10^{19}$ cm$^{-3}$ and $10^{21}$ cm$^{-3}$.\cite{Halliburton_Kappers_prl_1973,Halliburton_Cowan_prb_1973,Rose_Halliburton_jphys_1974,Araujo_Kapilashrami_apl_2010} 
According to density functional calculations, \cite{Araujo_Kapilashrami_apl_2010}  the magnetic moment of these vacancies is $m_{V_{Mg}}\approx 1.9\mu_B$. 
Whereas the number of vacancies might be smaller than the density of spinfull Mg nuclei, their magnetic moment is also 2000 times larger.  Thus,  they could also be the source of  more spin noise.   The analysis of the numerical data shows that, in both cases, 
$\sigma_{ {\cal  B}_a}\propto \sqrt{\langle m^2\rangle} \sqrt{\rho}/(wR)$ for $R\gg d,w$, 
so the standard deviation of the field scales with the square root of the barrier spin density, $\rho$.

We now consider the spectral properties of the nuclear and vacancy magnetic field noise. 
For that matter, we assume that every nuclear and vacancy spin precess freely under the influence of the magnetic field created by the ferromagnetic electrodes, $\vec{B}_{ext}$. Thus, we neglect the mutual coupling between spin centers in the barrier, except for a phenomenological relaxation time $T_1$ explained below.
Notice that the precession frequency of nuclear and electronic spins is very different, on account of their different magnetic moment. Then, for a MgO average field of 0.1 T, the nuclear and electronic precession frequencies are in the range of MHz and GHz respectively.

We assume that the magnetic field felt by the barrier spins is time independent and it only varies in the direction perpendicular to the interfaces ($z$).  This  approximation works well as long as the time fluctuations of the magnetic field created by the barrier are slow compared to the barrier spin dynamics.
Under these approximations, the correlation function for the detector average  at different times,
$S^2_a(t)\equiv \langle{\cal B}_a(t) {\cal B}_a(0)\rangle$, with $t>0$, is related to 
the spin correlation functions as
\beqa
S^2_a(t)= \sum_{ii',bb'}\Xi_{ab}(i)\Xi_{ab'}(i')\langle  m_b(i;t)m_{b'}(i';0)\rangle.
&& \label{correla2a}
\eeqa
The evaluation of this quantity is greatly simplified using the fact that, to a very good approximation,   different  barrier spins are uncorrelated.   Accordingly, the experimentally relevant  noise spectrum,  $S_x^2(\omega)=\int_{-\infty}^{\infty} e^{-i\omega t} S^2_x(t) dt$, can be expressed as:
\beqa
S^2_x(\omega)= \sum_{i,bb'}\Xi_{xb}(i)\Xi_{xb'}(i)
\langle m_b(i)m_{b'}(i)\rangle[\omega].
&& \label{correla2b}
\eeqa

 If we quantize the system along the magnetic field orientation at each nuclear spin, and denoting as $|n\rangle $ the nuclear spin eigenstates, the barrier spin spectral function reads,
  in the limit $k_B T\ggg  |\vec{m}|B_{ext}$,
\beqa
\langle m_b(i)m_{b'}(i)\rangle[\omega]&=& \frac{\delta_{i,i'}}{(2I+1)}
\sum_{nn'} \langle n|m_b|n'\rangle 
\crcr
&&\hspace{-1.cm}\times
 \langle n'|m_{b'}|n\rangle \delta(\omega-\omega_{nn'}(i)),
\label{corrm2}
\eeqa
where $\hbar \omega_{nn'}(i)= |\vec{m}|B_{ext}(i)(n-n')$ is the energy of the spin transition $n\to n'$, which depends on local the value of the external field. 
Some  straightforward algebra permits  obtaining the following relation between the spectral noise response $S_x(\omega)$ and $\sigma_{{\cal B}_x}$
\beqa
\int_{-\infty}^{\infty} S_x^2(\omega)d\omega=\frac{\sigma_{{\cal B}_x}^2}{3}.
\label{newS}
\eeqa

As a first approach, let us assume that all the barrier spins feel the same magnetic field intensity. 
Then, the $^{25}$Mg nuclear spins spectral function has a single finite-frequency peak at the Larmor frequency $\omega_B=|\vec{m}|B_{ext}/\hbar$.

\begin{figure}
\includegraphics[width=0.95\linewidth,angle=0]{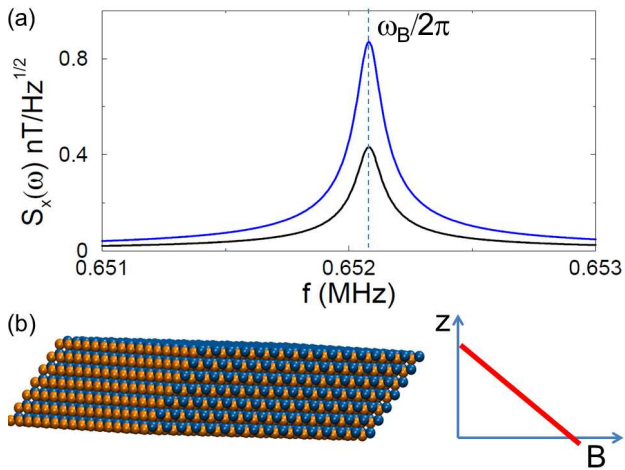}
\caption{ \label{fig2} (a) Spectral response $S_x(\omega)$ versus frequency $f=\omega/2\pi$
 for a detector of radius $R=100$ nm (black line) and $R=50$ nm (blue line), $d=1$ nm, $w=3$ nm, $B_{ext}=0.1$ T and $T_1=10$ ms. 
b) Scheme of the variation of the field along a 1 nm thick MgO layer.
}
\end{figure}

Due to its coupling to the environment,  the spectral function of a single nuclear spin, Eq. ({\ref{corrm2}), acquires a finite linewidth. We model this  by substituting the delta function in Eq. (\ref{corrm2}) by a Lorentzian function with a width $\delta \omega=2\pi/T_1$, with $T_1$ the  characteristic relaxation time. Typically, $T_1\lesssim 50 $ s in bulk MgO at room temperature,\cite{Fiske_Stebbins_physchem_1994} and it is expected to be at least $1$ms or larger in surfaces.\cite{Freitas_Smith_arep_2012} 
The resulting nuclear noise spectrum is shown in Fig. 2  for two values of $R$.  The magnitude of the peak noise associated to the nuclear spins is in the range of nT/Hz$^{1/2}$, centered in the Larmor frequency ($~0.5$MHz for $B_{ext}\sim 0.1$ T).

This reported nuclear noise has to be compared with the noise coming from other sources, such as
the $1/f$ noise.  We take as a reference a $R=20\mu$m sensor that has a noise level of pT/$\sqrt{{\rm Hz}}$ at 500 KHz.\cite{Chaves_Freitas_apl_2007,Chaves_Freitas_apl_2008}  We use the fact that the $1/f$ noise also scales like $1/R$ with size, so that,  extrapolating down to $R=100nm$, the $1/f$ noise  would be 0.4 nT/$\sqrt{{\rm Hz}}$,  comparable  to the one in Fig. 2(a).  Therefore the contributions of nuclear spin noise and $1/f$ noise are, under these assumptions, of the same order.

We now consider the noise due to spinful Mg vacancies. If we assume a lower limit for the $V_{Mg}$ concentration of $10^{19}$cm$^{-3}$, a small MgO layer of $R=25$ nm and $d=0.5nm$ will contain more than 10 vacancies.  $10^{4}$ $V_{Mg}$. 
Since the magnetic moment of these vacancies is around $1.9\mu_B$, at least three orders of magnitude larger than in the $^{25}$Mg nuclei,
even a single vacancy can produce fluctuations of the magnetic field of the order of $\mu$T for devices with $R=100$ nm, see inset of Fig. \ref{fig3}. A second consequence of the large difference in magnetic moment with the nuclei is that the corresponding 
Larmor frequency for typical fields around $0.1T$ will be in the range of GHz.

The magnitude of the field, which determines the location of the spectral noise peak,  is expected to change  along the MgO layer since, in general, the magnetization on the FL and pinning layer is different. 
 Magnetic field gradients up to $40$ mT/nm have been reported for magnetic disk heads.\cite{Tsang_Bonhote_ieee_2006}
In Fig. 3 we show the effect of a magnetic field gradient of 1 mT/nm.  Expectedly, several peaks appear in the spectrum corresponding to different Larmor frequencies, whose position reflects variations of the field across the different Mg atomic planes, see Fig.~\ref{fig2}(b).

\begin{figure}
\includegraphics[width=0.95\linewidth,angle=0]{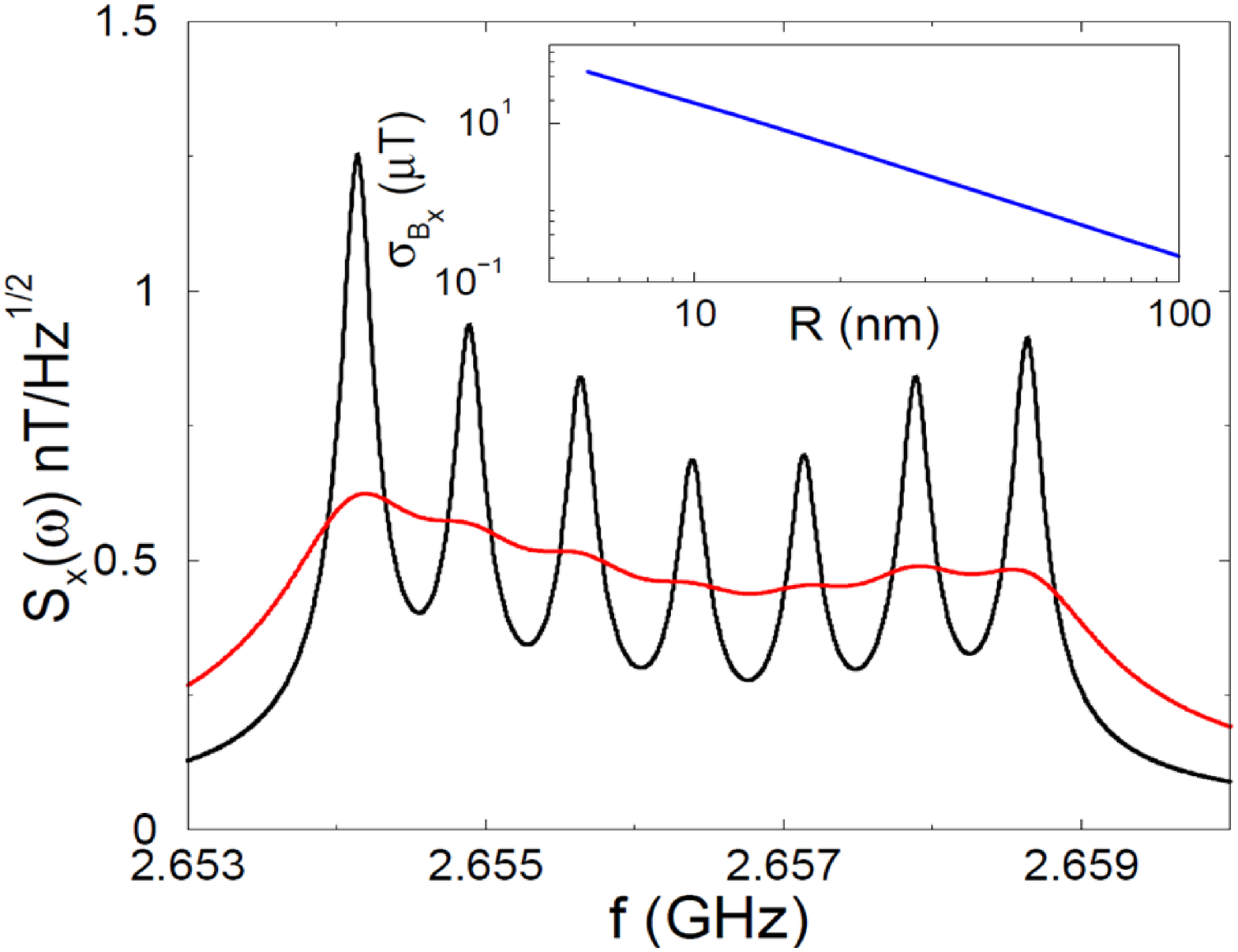}
\caption{ \label{fig3} Spectral response $\sigma(\omega)$ versus frequency $f=\omega/2\pi$
 for a detector of radius $R=100$ nm, $d=1$ nm, $w=3$ nm and  $T_1=5\mu$s (black line) and $T_1=1\mu$s (red line), containing 320 V$_{\rm Mg}$. 
A magnetic field gradient of $1$mT/nm along the $z$-axis was assumed. 
Inset shows the integrated standard deviation $\sigma_{{\cal B}_x}$ due to a single $V_{Mg}$ located at the center of the MgO layer versus the radius $R$.
}
\end{figure}

The different peaks will be resolved if their spectral broadening is smaller than the splitting, $|\vec{m}|.|\partial_z B(z)|d/\hbar\gg 2\pi/T_1$.
The relaxation time of these vacancies is much shorter than for the nuclear spins, below 100$\mu$s.\cite{Ferrari_Pacchioni_jphyschem_1995}
Figure \ref{fig3} shows the spectra corresponding to two different relaxation times, $T_1=$1 and 5$\mu$s. 
In both cases,  the relative height of the different peaks will reflect the abundance of vacancies in each atomic plane of the MgO. Thereby,  structural information concerning the distribution of Mg vacancies along the barrier could be inferred from measurements of the noise spectrum.

In conclusion, we have studied the impact of the fluctuating magnetic field created both by the $^{25}$Mg nuclear spins and Mg vacancies on a TMR magnetic field sensor with a thin MgO barrier, with circular section of radius $R$. 
The noise 
decreases inversely proportional to $R$ and it is spectrally peaked at the spin Larmor frequency, determined by the magnetic field 
in the barriers, which is typically in the range of 500 kHz for the nuclear spins and $2$GHz for the Mg vacancies. We argue that although 
the nuclear-induced noise in the 0.5MHz region is around 1 nT/$\sqrt{{\rm Hz}}$ for devices with $R=100$ nm, comparable to the $1/f$ noise,
the vacancies-induced noise should be larger than 1 nT/$\sqrt{{\rm Hz}}$  in the $2$ GHz vicinity, well above the $1/f$ noise.
We show that for a linearly varying magnetic field in the barrier, the noise spectrum can show a series of peaks whose position  and height reflects the  variations of the magnetic field magnitude and barrier spin density at the different Mg planes.   Thus, measurement of this noise, through electrical characterization, could provide some sort of spin imaging of the barrier.

  We acknowledge C. Untied for fruitful discussions. This work has been financially supported by MEC-Spain (Grant Nos. FIS2010-21883-C02-01, FIS2009-08744,  and CONSOLIDER CSD2007-0010) as well as Generalitat Valenciana, grant Prometeo 2012-11.




\begin{thebibliography}{23}
\expandafter\ifx\csname natexlab\endcsname\relax\def\natexlab#1{#1}\fi
\expandafter\ifx\csname bibnamefont\endcsname\relax
  \def\bibnamefont#1{#1}\fi
\expandafter\ifx\csname bibfnamefont\endcsname\relax
  \def\bibfnamefont#1{#1}\fi
\expandafter\ifx\csname citenamefont\endcsname\relax
  \def\citenamefont#1{#1}\fi
\expandafter\ifx\csname url\endcsname\relax
  \def\url#1{\texttt{#1}}\fi
\expandafter\ifx\csname urlprefix\endcsname\relax\def\urlprefix{URL }\fi
\providecommand{\bibinfo}[2]{#2}
\providecommand{\eprint}[2][]{\url{#2}}

\bibitem[{\citenamefont{Parkin et~al.}(2004)\citenamefont{Parkin, Kaiser,
  Panchula, Rice, Hughes, Samant, and Yang}}]{Parkin_Kaiser_natmat_2004}
\bibinfo{author}{\bibfnamefont{S.}~\bibnamefont{Parkin}},
  \bibinfo{author}{\bibfnamefont{C.}~\bibnamefont{Kaiser}},
  \bibinfo{author}{\bibfnamefont{A.}~\bibnamefont{Panchula}},
  \bibinfo{author}{\bibfnamefont{P.}~\bibnamefont{Rice}},
  \bibinfo{author}{\bibfnamefont{B.}~\bibnamefont{Hughes}},
  \bibinfo{author}{\bibfnamefont{M.}~\bibnamefont{Samant}}, \bibnamefont{and}
  \bibinfo{author}{\bibfnamefont{S.}~\bibnamefont{Yang}},
  \bibinfo{journal}{Nature materials} \textbf{\bibinfo{volume}{3}},
  \bibinfo{pages}{862} (\bibinfo{year}{2004}).

\bibitem[{\citenamefont{Yuasa et~al.}(2004)\citenamefont{Yuasa, Nagahama,
  Fukushima, Suzuki, and Ando}}]{Yuasa_Nagahama_natmat_2004}
\bibinfo{author}{\bibfnamefont{S.}~\bibnamefont{Yuasa}},
  \bibinfo{author}{\bibfnamefont{T.}~\bibnamefont{Nagahama}},
  \bibinfo{author}{\bibfnamefont{A.}~\bibnamefont{Fukushima}},
  \bibinfo{author}{\bibfnamefont{Y.}~\bibnamefont{Suzuki}}, \bibnamefont{and}
  \bibinfo{author}{\bibfnamefont{K.}~\bibnamefont{Ando}},
  \bibinfo{journal}{Nature materials} \textbf{\bibinfo{volume}{3}},
  \bibinfo{pages}{868} (\bibinfo{year}{2004}).

\bibitem[{\citenamefont{Ingvarsson et~al.}(2000)\citenamefont{Ingvarsson, Xiao,
  Parkin, Gallagher, Grinstein, and Koch}}]{Ingvarsson_Xiao_prl_2000}
\bibinfo{author}{\bibfnamefont{S.}~\bibnamefont{Ingvarsson}},
  \bibinfo{author}{\bibfnamefont{G.}~\bibnamefont{Xiao}},
  \bibinfo{author}{\bibfnamefont{S.~S.~P.} \bibnamefont{Parkin}},
  \bibinfo{author}{\bibfnamefont{W.~J.} \bibnamefont{Gallagher}},
  \bibinfo{author}{\bibfnamefont{G.}~\bibnamefont{Grinstein}},
  \bibnamefont{and} \bibinfo{author}{\bibfnamefont{R.~H.} \bibnamefont{Koch}},
  \bibinfo{journal}{Phys. Rev. Lett.} \textbf{\bibinfo{volume}{85}},
  \bibinfo{pages}{3289} (\bibinfo{year}{2000}).

\bibitem[{\citenamefont{Parkin et~al.}(2003)\citenamefont{Parkin, Jiang,
  Kaiser, Panchula, Roche, and Samant}}]{Parkin_Jiang_ieee_2003}
\bibinfo{author}{\bibfnamefont{S.}~\bibnamefont{Parkin}},
  \bibinfo{author}{\bibfnamefont{X.}~\bibnamefont{Jiang}},
  \bibinfo{author}{\bibfnamefont{C.}~\bibnamefont{Kaiser}},
  \bibinfo{author}{\bibfnamefont{A.}~\bibnamefont{Panchula}},
  \bibinfo{author}{\bibfnamefont{K.}~\bibnamefont{Roche}}, \bibnamefont{and}
  \bibinfo{author}{\bibfnamefont{M.}~\bibnamefont{Samant}},
  \bibinfo{journal}{Proceedings of the IEEE} \textbf{\bibinfo{volume}{91}},
  \bibinfo{pages}{661} (\bibinfo{year}{2003}).

\bibitem[{\citenamefont{Klaassen et~al.}(2005)\citenamefont{Klaassen, Xing, and
  van Peppen}}]{Klaassen_Xing_ieee_2005}
\bibinfo{author}{\bibfnamefont{K.}~\bibnamefont{Klaassen}},
  \bibinfo{author}{\bibfnamefont{X.}~\bibnamefont{Xing}}, \bibnamefont{and}
  \bibinfo{author}{\bibfnamefont{J.}~\bibnamefont{van Peppen}},
  \bibinfo{journal}{Magnetics, IEEE Transactions on}
  \textbf{\bibinfo{volume}{41}}, \bibinfo{pages}{2307} (\bibinfo{year}{2005}).

\bibitem[{\citenamefont{Freitas et~al.}(2007)\citenamefont{Freitas, Ferreira,
  Cardoso, and Cardoso}}]{Freitas_Ferreira_jphysc_2007}
\bibinfo{author}{\bibfnamefont{P.}~\bibnamefont{Freitas}},
  \bibinfo{author}{\bibfnamefont{R.}~\bibnamefont{Ferreira}},
  \bibinfo{author}{\bibfnamefont{S.}~\bibnamefont{Cardoso}}, \bibnamefont{and}
  \bibinfo{author}{\bibfnamefont{F.}~\bibnamefont{Cardoso}},
  \bibinfo{journal}{Journal of Physics: Condensed Matter}
  \textbf{\bibinfo{volume}{19}}, \bibinfo{pages}{165221}
  (\bibinfo{year}{2007}).

\bibitem[{\citenamefont{Egelhoff et~al.}(2009)\citenamefont{Egelhoff, Pong,
  Unguris, McMichael, Nowak, Edelstein, Burnette, and
  Fischer}}]{Egelhoff_Pong_sensors_2009}
\bibinfo{author}{\bibfnamefont{W.}~\bibnamefont{Egelhoff}},
  \bibinfo{author}{\bibfnamefont{P.}~\bibnamefont{Pong}},
  \bibinfo{author}{\bibfnamefont{J.}~\bibnamefont{Unguris}},
  \bibinfo{author}{\bibfnamefont{R.}~\bibnamefont{McMichael}},
  \bibinfo{author}{\bibfnamefont{E.}~\bibnamefont{Nowak}},
  \bibinfo{author}{\bibfnamefont{A.}~\bibnamefont{Edelstein}},
  \bibinfo{author}{\bibfnamefont{J.}~\bibnamefont{Burnette}}, \bibnamefont{and}
  \bibinfo{author}{\bibfnamefont{G.}~\bibnamefont{Fischer}},
  \bibinfo{journal}{Sensors and Actuators A: Physical}
  \textbf{\bibinfo{volume}{155}}, \bibinfo{pages}{217} (\bibinfo{year}{2009}).

\bibitem[{\citenamefont{Lei et~al.}(2011)\citenamefont{Lei, Li, Egelhoff, Lai,
  and Pong}}]{Lei_Li_magnetics_2011}
\bibinfo{author}{\bibfnamefont{Z.}~\bibnamefont{Lei}},
  \bibinfo{author}{\bibfnamefont{G.}~\bibnamefont{Li}},
  \bibinfo{author}{\bibfnamefont{W.}~\bibnamefont{Egelhoff}},
  \bibinfo{author}{\bibfnamefont{P.}~\bibnamefont{Lai}}, \bibnamefont{and}
  \bibinfo{author}{\bibfnamefont{P.}~\bibnamefont{Pong}},
  \bibinfo{journal}{Magnetics, IEEE Transactions on}
  \textbf{\bibinfo{volume}{47}}, \bibinfo{pages}{602} (\bibinfo{year}{2011}).

\bibitem[{\citenamefont{Berglund and
  Wieser}(2011)}]{Berglund_Wieser_pachem_2011}
\bibinfo{author}{\bibfnamefont{M.}~\bibnamefont{Berglund}} \bibnamefont{and}
  \bibinfo{author}{\bibfnamefont{M.~E.} \bibnamefont{Wieser}},
  \bibinfo{journal}{Pure and Applied Chemistry} \textbf{\bibinfo{volume}{83}},
  \bibinfo{pages}{397} (\bibinfo{year}{2011}).

\bibitem[{\citenamefont{Halliburton
  et~al.}(1973{\natexlab{a}})\citenamefont{Halliburton, Kappers, Cowan,
  Dravnieks, and Wertz}}]{Halliburton_Kappers_prl_1973}
\bibinfo{author}{\bibfnamefont{L.~E.} \bibnamefont{Halliburton}},
  \bibinfo{author}{\bibfnamefont{L.~A.} \bibnamefont{Kappers}},
  \bibinfo{author}{\bibfnamefont{D.~L.} \bibnamefont{Cowan}},
  \bibinfo{author}{\bibfnamefont{F.}~\bibnamefont{Dravnieks}},
  \bibnamefont{and} \bibinfo{author}{\bibfnamefont{J.~E.} \bibnamefont{Wertz}},
  \bibinfo{journal}{Phys. Rev. Lett.} \textbf{\bibinfo{volume}{30}},
  \bibinfo{pages}{607} (\bibinfo{year}{1973}{\natexlab{a}}).

\bibitem[{\citenamefont{Halliburton
  et~al.}(1973{\natexlab{b}})\citenamefont{Halliburton, Cowan, Blake, and
  Wertz}}]{Halliburton_Cowan_prb_1973}
\bibinfo{author}{\bibfnamefont{L.~E.} \bibnamefont{Halliburton}},
  \bibinfo{author}{\bibfnamefont{D.~L.} \bibnamefont{Cowan}},
  \bibinfo{author}{\bibfnamefont{W.~B.~J.} \bibnamefont{Blake}},
  \bibnamefont{and} \bibinfo{author}{\bibfnamefont{J.~E.} \bibnamefont{Wertz}},
  \bibinfo{journal}{Phys. Rev. B} \textbf{\bibinfo{volume}{8}},
  \bibinfo{pages}{1610} (\bibinfo{year}{1973}{\natexlab{b}}).

\bibitem[{\citenamefont{Rose and
  Halliburton}(1974)}]{Rose_Halliburton_jphys_1974}
\bibinfo{author}{\bibfnamefont{B.}~\bibnamefont{Rose}} \bibnamefont{and}
  \bibinfo{author}{\bibfnamefont{L.}~\bibnamefont{Halliburton}},
  \bibinfo{journal}{Journal of Physics C: Solid State Physics}
  \textbf{\bibinfo{volume}{7}}, \bibinfo{pages}{3981} (\bibinfo{year}{1974}).

\bibitem[{\citenamefont{Araujo et~al.}(2010)\citenamefont{Araujo, Kapilashrami,
  Jun, Jayakumar, Nagar, Wu, Arhammar, Johansson, Belova, Ahuja
  et~al.}}]{Araujo_Kapilashrami_apl_2010}
\bibinfo{author}{\bibfnamefont{C.}~\bibnamefont{Araujo}},
  \bibinfo{author}{\bibfnamefont{M.}~\bibnamefont{Kapilashrami}},
  \bibinfo{author}{\bibfnamefont{X.}~\bibnamefont{Jun}},
  \bibinfo{author}{\bibfnamefont{O.}~\bibnamefont{Jayakumar}},
  \bibinfo{author}{\bibfnamefont{S.}~\bibnamefont{Nagar}},
  \bibinfo{author}{\bibfnamefont{Y.}~\bibnamefont{Wu}},
  \bibinfo{author}{\bibfnamefont{C.}~\bibnamefont{Arhammar}},
  \bibinfo{author}{\bibfnamefont{B.}~\bibnamefont{Johansson}},
  \bibinfo{author}{\bibfnamefont{L.}~\bibnamefont{Belova}},
  \bibinfo{author}{\bibfnamefont{R.}~\bibnamefont{Ahuja}},
  \bibnamefont{et~al.}, \bibinfo{journal}{Applied Physics Letters}
  \textbf{\bibinfo{volume}{96}}, \bibinfo{pages}{232505}
  (\bibinfo{year}{2010}).

\bibitem[{\citenamefont{Stone}(2005)}]{Stone_atdat_2005}
\bibinfo{author}{\bibfnamefont{N.~J.} \bibnamefont{Stone}},
  \bibinfo{journal}{Atomic Data and Nuclear Data Tables}
  \textbf{\bibinfo{volume}{90}}, \bibinfo{pages}{75} (\bibinfo{year}{2005}).

\bibitem[{\citenamefont{Sleator et~al.}(1985)\citenamefont{Sleator, Hahn,
  Hilbert, and Clarke}}]{Sleator_Hahn_prl_1985}
\bibinfo{author}{\bibfnamefont{T.}~\bibnamefont{Sleator}},
  \bibinfo{author}{\bibfnamefont{E.~L.} \bibnamefont{Hahn}},
  \bibinfo{author}{\bibfnamefont{C.}~\bibnamefont{Hilbert}}, \bibnamefont{and}
  \bibinfo{author}{\bibfnamefont{J.}~\bibnamefont{Clarke}},
  \bibinfo{journal}{Phys. Rev. Lett.} \textbf{\bibinfo{volume}{55}},
  \bibinfo{pages}{1742} (\bibinfo{year}{1985}).

\bibitem[{\citenamefont{Degen et~al.}(2007)\citenamefont{Degen, Poggio, Mamin,
  and Rugar}}]{Degen_Poggio_prl_2007}
\bibinfo{author}{\bibfnamefont{C.~L.} \bibnamefont{Degen}},
  \bibinfo{author}{\bibfnamefont{M.}~\bibnamefont{Poggio}},
  \bibinfo{author}{\bibfnamefont{H.~J.} \bibnamefont{Mamin}}, \bibnamefont{and}
  \bibinfo{author}{\bibfnamefont{D.}~\bibnamefont{Rugar}},
  \bibinfo{journal}{Phys. Rev. Lett.} \textbf{\bibinfo{volume}{99}},
  \bibinfo{pages}{250601} (\bibinfo{year}{2007}).

\bibitem[{\citenamefont{Wertz et~al.}(1959)\citenamefont{Wertz, Auzins,
  Griffiths, and Orton}}]{Wertz_Auzins_dfs_1959}
\bibinfo{author}{\bibfnamefont{J.}~\bibnamefont{Wertz}},
  \bibinfo{author}{\bibfnamefont{P.}~\bibnamefont{Auzins}},
  \bibinfo{author}{\bibfnamefont{J.}~\bibnamefont{Griffiths}},
  \bibnamefont{and} \bibinfo{author}{\bibfnamefont{J.}~\bibnamefont{Orton}},
  \bibinfo{journal}{Discussions of the Faraday Society}
  \textbf{\bibinfo{volume}{28}}, \bibinfo{pages}{136} (\bibinfo{year}{1959}).

\bibitem[{\citenamefont{Fiske et~al.}(1994)\citenamefont{Fiske, Stebbins, and
  Farnan}}]{Fiske_Stebbins_physchem_1994}
\bibinfo{author}{\bibfnamefont{P.~S.} \bibnamefont{Fiske}},
  \bibinfo{author}{\bibfnamefont{J.~F.} \bibnamefont{Stebbins}},
  \bibnamefont{and} \bibinfo{author}{\bibfnamefont{I.}~\bibnamefont{Farnan}},
  \bibinfo{journal}{Physics and Chemistry of Minerals}
  \textbf{\bibinfo{volume}{20}}, \bibinfo{pages}{587} (\bibinfo{year}{1994}).

\bibitem[{\citenamefont{Freitas and Smith}(2012)}]{Freitas_Smith_arep_2012}
\bibinfo{author}{\bibfnamefont{J.}~\bibnamefont{Freitas}} \bibnamefont{and}
  \bibinfo{author}{\bibfnamefont{M.}~\bibnamefont{Smith}},
  \bibinfo{journal}{Annual Reports on NMR Spectroscopy} p.~\bibinfo{pages}{25}
  (\bibinfo{year}{2012}).

\bibitem[{\citenamefont{Chaves et~al.}(2007)\citenamefont{Chaves, Freitas,
  Ocker, and Maass}}]{Chaves_Freitas_apl_2007}
\bibinfo{author}{\bibfnamefont{R.}~\bibnamefont{Chaves}},
  \bibinfo{author}{\bibfnamefont{P.}~\bibnamefont{Freitas}},
  \bibinfo{author}{\bibfnamefont{B.}~\bibnamefont{Ocker}}, \bibnamefont{and}
  \bibinfo{author}{\bibfnamefont{W.}~\bibnamefont{Maass}},
  \bibinfo{journal}{Applied Physics Letters} \textbf{\bibinfo{volume}{91}},
  \bibinfo{pages}{102504} (\bibinfo{year}{2007}).

\bibitem[{\citenamefont{Chaves et~al.}(2008)\citenamefont{Chaves, Freitas,
  Ocker, and Maass}}]{Chaves_Freitas_apl_2008}
\bibinfo{author}{\bibfnamefont{R.}~\bibnamefont{Chaves}},
  \bibinfo{author}{\bibfnamefont{P.}~\bibnamefont{Freitas}},
  \bibinfo{author}{\bibfnamefont{B.}~\bibnamefont{Ocker}}, \bibnamefont{and}
  \bibinfo{author}{\bibfnamefont{W.}~\bibnamefont{Maass}},
  \bibinfo{journal}{Journal of Applied Physics} \textbf{\bibinfo{volume}{103}},
  \bibinfo{pages}{07E931} (\bibinfo{year}{2008}).

\bibitem[{\citenamefont{Tsang et~al.}(2006)\citenamefont{Tsang, Bonhote, Dai,
  Do, Knigge, Ikeda, Le, Lengsfield, Lille, Li
  et~al.}}]{Tsang_Bonhote_ieee_2006}
\bibinfo{author}{\bibfnamefont{C.}~\bibnamefont{Tsang}},
  \bibinfo{author}{\bibfnamefont{C.}~\bibnamefont{Bonhote}},
  \bibinfo{author}{\bibfnamefont{Q.}~\bibnamefont{Dai}},
  \bibinfo{author}{\bibfnamefont{H.}~\bibnamefont{Do}},
  \bibinfo{author}{\bibfnamefont{B.}~\bibnamefont{Knigge}},
  \bibinfo{author}{\bibfnamefont{Y.}~\bibnamefont{Ikeda}},
  \bibinfo{author}{\bibfnamefont{Q.}~\bibnamefont{Le}},
  \bibinfo{author}{\bibfnamefont{B.}~\bibnamefont{Lengsfield}},
  \bibinfo{author}{\bibfnamefont{J.}~\bibnamefont{Lille}},
  \bibinfo{author}{\bibfnamefont{J.}~\bibnamefont{Li}}, \bibnamefont{et~al.},
  \bibinfo{journal}{Magnetics, IEEE Transactions on}
  \textbf{\bibinfo{volume}{42}}, \bibinfo{pages}{145} (\bibinfo{year}{2006}).

\bibitem[{\citenamefont{Ferrari and
  Pacchioni}(1995)}]{Ferrari_Pacchioni_jphyschem_1995}
\bibinfo{author}{\bibfnamefont{A.}~\bibnamefont{Ferrari}} \bibnamefont{and}
  \bibinfo{author}{\bibfnamefont{G.}~\bibnamefont{Pacchioni}},
  \bibinfo{journal}{The Journal of Physical Chemistry}
  \textbf{\bibinfo{volume}{99}}, \bibinfo{pages}{17010} (\bibinfo{year}{1995}).

\end{thebibliography}
\end{document}